# Electro-optic frequency comb-enabled precise distance measurement with megahertz acquisition rate

Yifan Qi [1,3], Xingyu Jia [1,3], Jingyi Wang [2], Weiwei Yang [1], Yihan Miao [1], Xinlun Cai [2,*], Guanhao Wu [1,**] and Yang Li [1,***]

[1] State Key Laboratory of Precision Measurement Technology and Instruments, Department of Precision Instrument, Tsinghua University, Beijing, 100084, China

[2] State Key Laboratory of Optoelectronic Materials and Technologies, School of Electronics and Information Technology, Sun Yat-sen University, Guangzhou. 510275, China

[3] These authors contributed equally: Yifan Qi, Xingyu Jia.

\* *caixinlun5@mail.sysu.edu.cn*, \*\* *Guanhaowu@mail.tsinghua.edu.cn*, \*\*\* *yli9003@tsinghua.edu.cn*

## Abstract

Artificial intelligence empowered autonomous vehicles and robotics have to sense the fast-changing three-dimensional environment with high precision and speed. However, it is challenging for the state-of-the-art ambiguity-free light detection and ranging (LiDAR) techniques to achieve absolute distance measurement with simultaneous high precision and high acquisition rate. Here we demonstrate an electro-optic frequency comb-enabled precise absolute distance measurement method, repetition rate modulated frequency comb (RRMFC), with megahertz-level acquisition rate. To achieve RRMFC, we designed and fabricated an integrated lithium niobate phase modulator with a modulation length of 5 cm and a half-wave voltage of 1.52 V, leading to over 50 sidebands and a continuously tunable repetition rate. Leveraging these unique features, RRMFC can directly resolve distance in time domain, leading to an acquisition rate as high as 25 MHz and an Allan deviation down to 13.77 μm at an averaging time of 724 μs. Based on RRMFC, we achieved 3 megapixel/s 3D imaging at millimeter-level precision with a single laser. RRMFC-based LiDAR allows the autonomous vehicles and robotics to sense the fine details of fast-changing environment with high precision.

## Main

With the blooming of artificial intelligence technology, the distance sensing systems are highly demanded by autonomous vehicles, drones and robot products, virtual and augmented reality (VR and AR) systems[1-6]. Agile deployment, high acquisition rate, and precision are desirable for a sensing system to map the fast-changing environment by three dimensional (3D) point cloud information over various weather and illumination conditions. To date, mainstream distance sensing system for light detection and ranging (LiDAR) are direct time of flight (dToF) LiDAR and frequency modulation (frequency modulated continuous wave, FMCW) LiDAR[7-12]. FMCW, as a coherent detection method, shows high resolution, high selectivity against ambient light, innate capability for speed sensing and eye safety[13,14], leading to its fast emerging in many applications such as autonomous driving control.

FMCW requires Fourier transform to resolve the distance information[2,3], imposing severe limits on the acquisition rate of the system. The key components of FMCW LiDAR are frequency modulated laser and coherent receiver. The laser frequency of FMCW is chirped linearly (mostly with triangular waveform in frequency) to resolve the distance information from the frequency difference between the transmitted and received signals in one chirp period (fig. 1b)[15]. To resolve this frequency difference from the received time-domain sinusoidal signal, a discrete Fourier transform (typically fast Fourier transform, FFT) is performed, introducing limitations on integral time and distance resolution. The limitation on integral time decreases the acquisition rate, and also places higher demand on the span, speed, and linearity of frequency chirping[2,16-18]. This limitation is one of the main reasons that prevents FMCW from being used in several areas such as autonomous driving despite FMCW's numerous advantages in comparison to dToF.

Here we demonstrate an electro-optic comb-enabled absolute distance measurement method and system with megahertz acquisition rate. To achieve this system, repetition rate modulated frequency comb (RRMFC) coherent distance measurement method is proposed and is enabled by an integrated lithium niobate electro-optic frequency comb. The proposed coherent distance measurement system can achieve an acquisition rate up to 25 MHz at 23-m optical path and an Allan deviation down to 13.77 μm at 3 m optical length. This integrated lithium niobate frequency comb is designed and packaged for greater modulation depth, continuously adjustable repetition rate, and lower power consumption, which brings more frequency comb lines with commercially available continuous wave lasers and frequency-sweep microwave sources. In RRMFC ranging, all the comb lines are sent into a coherent ranging system, which carries out a temporal interference with high fineness (fig. 1c). We show that real-time distance resolving can be achieved with only two adjacent interference peaks among all peaks generated in one modulation frequency chirp period, boosting the acquisition rate of LiDAR system for ambiguity-free distance measurement.

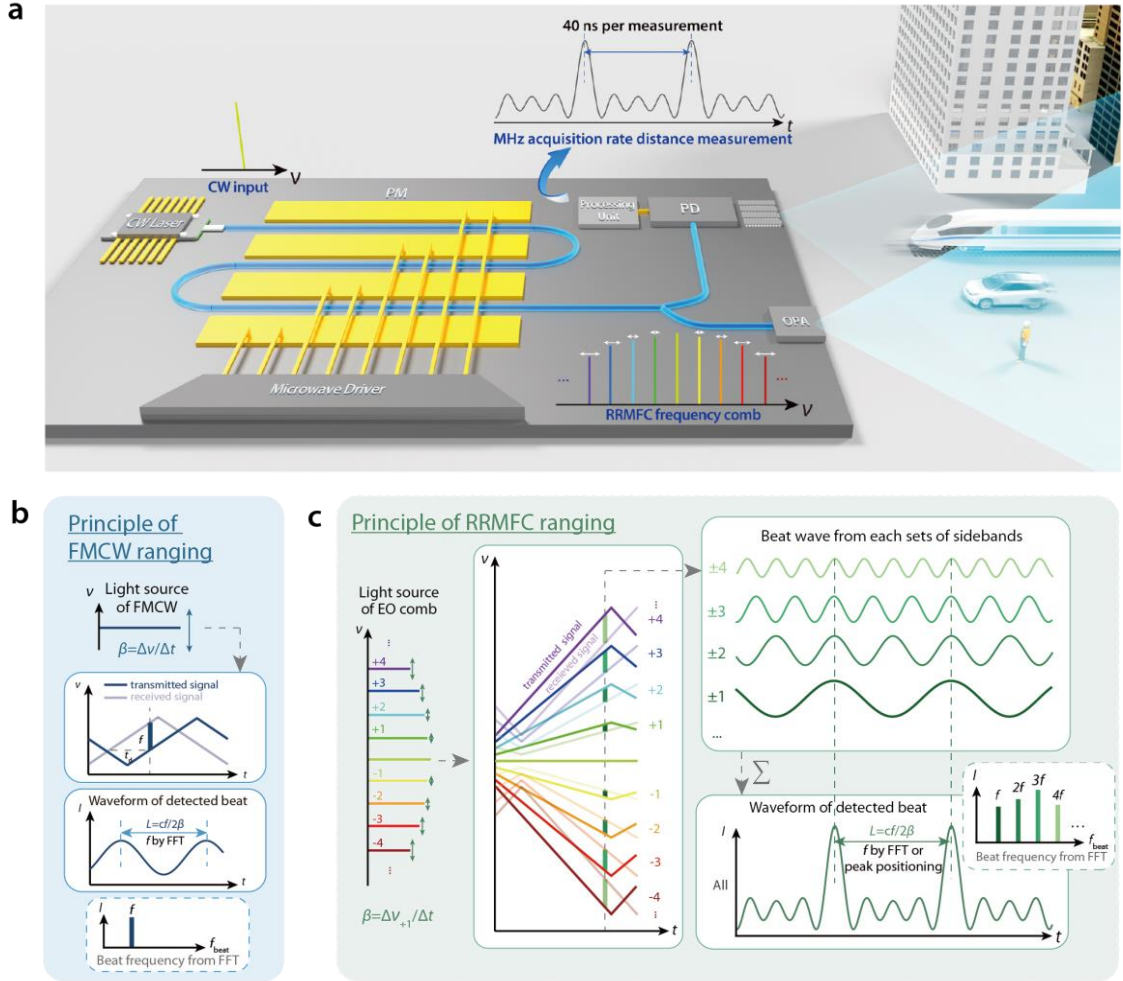

**Figure 1: Schematic and principles of frequency modulation coherent ranging.** (a) Schematic of RRMFC. PM: phase modulator. OPA: optical phase array. PD: photodiode. (b) Principle of FMCW. A triangular frequency modulated continuous wave with modulation rate $\beta$ is sent into coherent ranging system. The received signal and a part of the transmitted signal are detected together, forming a beat signal whose frequency $f$ is proportional to the time delay $t_d$ induced by the distance to be measured $L$. By getting $f$ in the beatnote spectrum, the distance can then be calculated via $cf/2\beta$. (c) Principle of RRMFC. Based on an electro-optic frequency comb with a tunable repetition rate, multiple sets of triangular frequency modulated continuous waves are generated by double sideband modulation. In the coherent ranging system, each set of double sidebands generates a beat signal with multiplied frequency and the same initial phase. The photodiode detects the sum of all the beat signals generated by all sets of double sidebands. By introducing higher-order beat signals with more modulation sidebands, the peaks of time-domain interferogram becomes sharper, enabling acquiring the beat frequency via two adjacent peaks and in turn a high acquisition rate. Alternatively, the beat frequency can also be acquired by FFT for a better immunity to ambient light.

## Principle of distance measurement

Time-domain RRMFC Lidar utilizes an electro-optic frequency comb with agile repetition rate adjustment as the light source for frequency modulated coherent ranging (fig. 1c). An electro-optic frequency comb can generate multiple sets of frequency modulated sidebands via double sideband modulation[19-22]. By tuning the repetition rate with a linearly chirped microwave signal, the frequency of each sideband will sweep linearly and synchronously. Due to the cumulative effect of repetition rate, the frequency sweeping speed of each sideband $n\beta$ is proportional to the sideband order $n$. As a result, the simultaneous and coherent detection of the transmitted and received signals of all sidebands generates a series of beat signals with beat frequency of $nf$ and the same phase. Accordingly, the series of beat frequencies correspond to a time-domain interferogram with the same period as FMCW but much sharper peaks. Such sharp peaks enable us to directly measure the beat frequency and the distance via $L=cf/2\beta$ by resolving the spacing between two adjacent peaks in time domain, resulting in an FFT-free data processing and in turn a much higher acquisition rate.

RRMFC shows the potential to achieve a much higher acquisition rate in frequency-modulated coherent ranging methods. Conventional frequency-modulated coherent ranging systems, such as FMCW, measure distance by resolving the beat frequency $f$ in the spectrum of time-domain interferogram obtained using FFT, necessitating a wide modulation frequency sweeping bandwidth[23,24] and a long integral time for a high resolution. In contrast, RRMFC resolves $f$ by measuring its corresponding time-domain period $1/f$ in the interferogram, leading to many effective distance measurements per modulation frequency sweeping period. Furthermore, considering a longer distance ($L=cf/2\beta$) results in a higher $f$ corresponding to a shorter time-domain period $1/f$, RRMFC can achieve a higher acquisition rate when measuring a longer distance.

In addition to the high acquisition rate, RRMFC also has the potential to achieve a high resolution and an infinite ambiguity range. Because RRMFC measures the distance by resolving the sharp peaks in time-domain interferogram, RRMFC's resolution is inversely proportional to the linewidth of the peaks. Hence, we can improve the resolution by reducing the linewidth of time-domain peaks through increasing the bandwidth of electro-optic frequency comb. Furthermore, according to distance resolving equation $L=cf/2\beta$, RRMFC's ambiguity range is limited by the highest beat frequency $f$, corresponding to the shortest period $1/f$ of time-domain interferogram. This shortest period can be measured as long as it is shorter than the chirp period of the modulation signal. Hence, RRMFC's ambiguity range is, in principle, only limited by the coherence length and power of the laser because it's trivial to sweep the modulation signal over a small frequency range.

**Device for distance measurement**

We generated the electro-optic frequency comb via an integrated lithium niobate phase modulator. Phase modulator tunes the instantaneous phase of the optical wave according to a microwave electric signal[25] (fig. 2A). In the frequency domain, the output light behaves as an optical frequency comb whose sidebands' intensities show a distribution according to the Bessel function. When we linearly sweep the frequency of the microwave modulation signal of a phase modulator, the spacing between adjacent sidebands (repetition rate, $f_r$) is locked to the modulation frequency ($f_m$) via the intrinsic electro-optic effect of lithium niobate. And, the quantity and intensity of sidebands are determined by the power of modulation signal and the design of phase modulator.

We designed and fabricated an integrated lithium niobate frequency comb based on a folded phase modulator with a modulation length up to 5 cm, leading to more sidebands for RRMFC at a moderate modulation power. Compared to the conventional design of integrated lithium niobate modulator with ridge waveguide and segmented electrodes, we used a folded waveguide and electrode configuration to extend the interaction length of electro-optic modulation with two 180° turning, leading to more sidebands within a limited footprint (fig. 2b). In each turning area, we designed waveguide crossing to guarantee the consistent phase accumulation direction before and after turning[26]. We designed the turning waveguide and electrode with minimum length for a lower power consumption across the sweeping range of the modulation frequency. To minimize the folded electrodes-induced discontinuities and imbalance to the microwave signal, we implemented air bridge structures and chamfered bending signal electrodes[27-29] in the turning area (fig. 2b).

We measured the electric and optical performance of the integrated lithium niobate frequency comb. The measured half-wave voltage of our device is 1.47 V at 3 GHz and 1.85 V at 9.5 GHz without push-pull electrodes structure, which outperform state-of-the-art non-resonant integrated lithium niobate modulator operating at 1550 nm wavelength[25]. Furthermore, we packaged our chip for stability and heat dissipation (fig. 2d), allowing for a high-power driving voltage and in turn more sidebands (fig. 2e). Our modulator can generate 59 sidebands with 28.5 dBm driving power, providing up to 29 sets of frequency modulated continuous waves for RRMFC. The combination of these sidebands yields a set of sharp peaks in the time-domain interferogram, enabling distance measurement by resolving the spacing between two adjacent peaks.

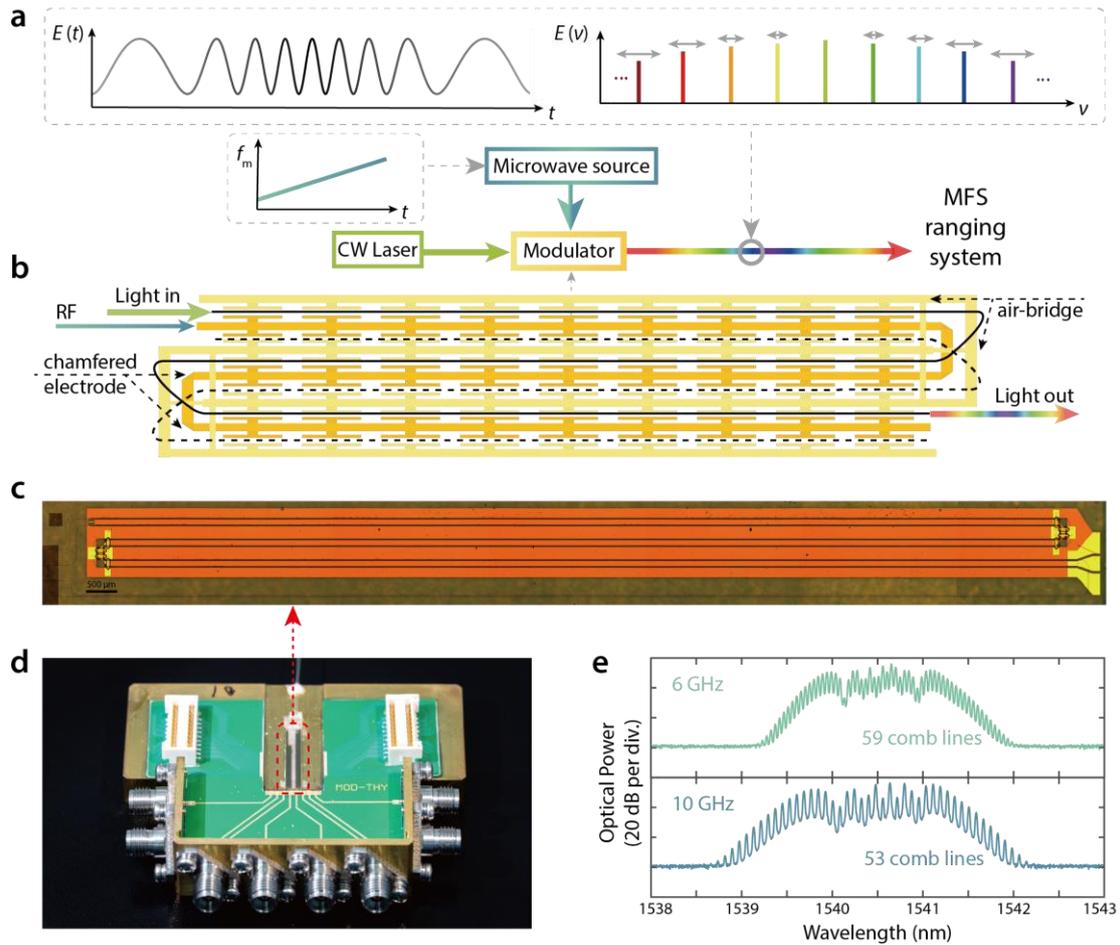

**Figure 2: Schematic, microscope image, photo, and measurement results of integrated lithium niobate frequency comb based on folded phase modulator.** (a) Schematic of the electro-optic phase modulator for generating an optical frequency comb. (b) Top-view schematic of the integrated lithium niobate folded phase modulator. (c) Microscope image of fabricated integrated lithium niobate folded phase modulator. (d) Photo of modulator after packaging. The red dashed box shows the integrated modulator in (c). (e) Optical frequency combs generated by the modulator.

# High-speed precise distance measurement

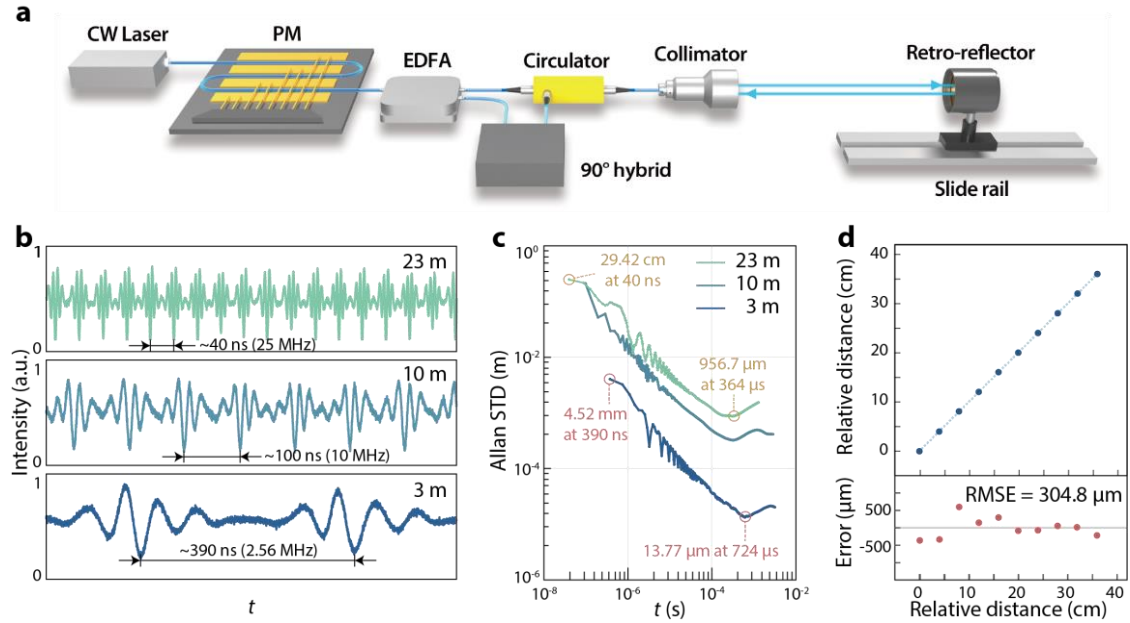

**Figure 3: Schematic and results of high-speed precise distance measurement using RRMFC.** (a) Experimental setup. The frequency comb is amplified by erbium-doped fiber amplifier (EDFA) and be split into measured arm and local oscillation arm. The optical signal in measured arm is sent to free space and reflected by a retro-reflector on a programmable slide rail, and is coherently detected by a 90° optical hybrid. (b) Measured time-domain interferograms under distances of 3 m, 10 m, and 23 m. Labels show the measured time spacing between two adjacent peaks and the corresponding acquisition rate. (c) Allan deviations of different measured distances as a function of averaging time. (d) (Top) Comparison of the measured distance versus the reference. (Bottom) Error of measured distance relative to reference. The target is placed on a motion stage and is measured by RRMFC and grating ruler simultaneously. "Target position" is the absolute distance of the target measured by RRMFC while "Relative distance" is the result measured by grating ruler.

We performed proof-of-concept absolute distance measurement experiments based on our RRMFC system using time-domain peak positioning. The experiment setup is depicted in fig. 3a. Electro-optic comb with repetition rate sweeping from 6.5 GHz to 11.5 GHz within 4 μs is generated by an integrated lithium niobate folded phase modulator and is sent into a distance measurement channel and a local oscillation channel. The time-domain interferograms (fig. 3b) show periodic sharp peaks, enabling resolving the instantaneous beat frequency and in turn the instantaneous distance from the time difference between two adjacent peaks. The measurement refresh time is 40 ns under a 23-m optical path difference, corresponding to an ultra-high distance acquisition rate up to 25 MHz. The acquisition rate decreases as the distance decreases, as predicted by the principle of RRMFC. The acquisition rate can be further improved by using a microwave source with a faster frequency-sweeping speed and photodiodes with wider bandwidth.

To test the precision and stability of RRMFC, we evaluated the Allan deviations of RRMFC distance measurement under distances of 3 m, 10 m, and 23 m. As shown in fig. 3c, a longer distance shows the Allan deviation starting at a shorter averaging time because a longer distance corresponds to a shorter period in time-domain interferogram and in turn a higher acquisition rate. Furthermore, for a given averaging time, a longer distance yields a larger Allan deviation due to following reasons. First, for a certain ratio between the linewidth of peak and period of time-domain interferogram, ranging error is proportional to distance. Second, for a given time-domain resolution of data acquisition hardware, the shorter period of time-domain interferogram corresponding to a longer distance induces a larger relative error. Third, the nonlinear error of the microwave frequency sweeping induces a certain relative error to the measured period of time-domain interferogram, leading to a larger absolute error for a smaller time-domain period corresponding to a longer distance. For each distance, as averaging time increases, Allan deviation first decreases and then increases, showing the maximum of 29.42 cm at the averaging time of 0.04 μs (corresponding to acquisition rate of 25 MHz) under a 23-m distance and the minimum of 13 μm at the averaging time of 724 μs (corresponding to acquisition rate of 1.4 kHz) under a 3-m distance. Allan deviation increases after a certain averaging time because of the environmental noises such as vibration of fibers.

In addition to precision and stability, we also tested the accuracy of RRMFC. Fig. 3d shows the accuracy of measured distance as a function of optical path difference starting from 14.9458 m. This optical path difference is achieved by moving the target using a motion stage monitored by a grating ruler with positioning accuracy better than 10 μm. The measurement accuracy (defined as the standard deviation (STD) of the residual errors) can reach 304.8 μm (relative accuracy $2.042 \times 10^{-5}$). This accuracy can be further improved by reducing the nonlinearity of microwave source, vibrations of the setup, and environmental conditions, such as the fiber forming the optical path.

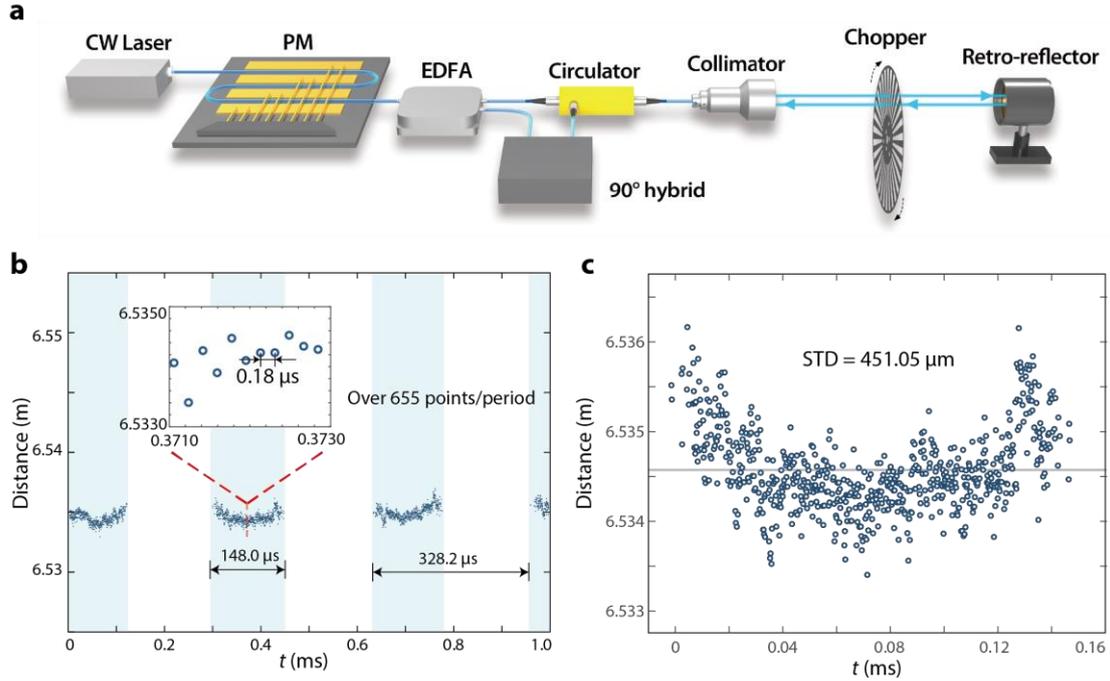

**Figure 4: Schematic and results of ultra-high acquisition rate test.** (a) Experimental setup. The output light in measured arm is chopped by an optical chopper. (b) Measured surface profile of the chopper. Inset shows the time spacing between two adjacent distance measurements. Blue dot shows the measured distance from the retroreflector. (c) Measured distance and precision within one chopping period.

We further tested RRMFC's high acquisition rate and precision by measuring the distance through a fast-rotating chopper. As shown in fig. 4a, we placed a 30-slot chopper with a chopping frequency of 3 kHz between the circulator and retro-reflector. When the repetition rate of the electro-optic comb is sweeping over the range of 1 GHz at sweeping frequency of 300 kHz, more than 655 measurements can be performed within one chopping period around 330 μs (fig. 4b), demonstrating RRMFC's capability in profile measurement of a fast moving object. As shown in the inset of fig. 4b, the typical measurement time is 0.18 μs, corresponding to an acquisition rate of 5.56 MHz. This acquisition rate can be further improved with a longer optical path and a higher optical transmit power. Fig. 4b shows the precision of distance measurement within one chopping period. Here, we achieved a STD of 451.05 μm with a mean of 6.5346 m, corresponding to a relative uncertainty of $6.9\times10^{-5}$.

# High-speed 3D imaging

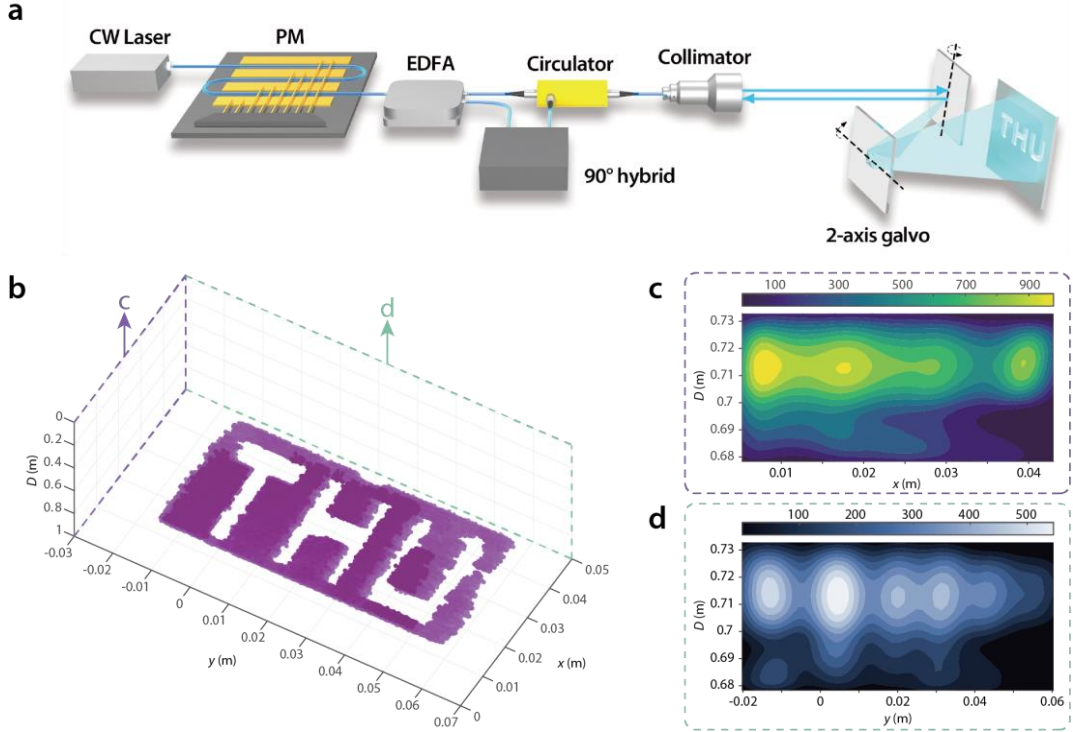

**Figure 5: Schematic and results of high-speed 3D imaging.** (a) Experimental setup. The output frequency comb signal is steered by a dual-axis galvanometer scanner. The target is an unpolished aluminum plate with 'THU' shaped holes. (b) 3D image reconstructed by the distance measurement results. (d, e) Projections of the 3D image along $x$ and $y$ axes. The colormap represents the superposition of the probability density functions of all points forming the point cloud.

The high acquisition rate and precision of RRMFC makes it an appealing solution to 3D imaging with a single laser. We performed a proof-of-concept demonstration of high-speed 3D imaging based on RRMFC system. As depicted in fig. 5a, we used a 2-axis scanning galvanometer to steer the light beam to different points along horizontal and vertical directions using progressive scan method. The light beam is reflected back from a metal sheet target with unpolished surface and engraved letters "THU". The target measurement result is shown in fig. 5b. The output light incident onto the flat area of the metal target is reflected back and coherently detected, generating the point cloud via the depth information provided by RRMFC. Benefiting from the high distance-resolving speed enabled by the time-domain peak positioning method, the point rate of this 3D imaging system reached 3 megapixel/s with a single laser. The point rate could be further increased via parallel ranging[9] by replacing the CW laser with a high repetition rate frequency comb. Fig. 5c and fig. 5d show the high density of the point cloud as well as the high distance measurement precision with a STD of 9.3 mm.

# Discussion

In summary, we demonstrated an electro-optic comb-enabled high-speed absolute distance measurement method and system. To achieve RRMFC, we designed and fabricated an integrated lithium niobate folded electro-optic frequency comb with high modulation depth and continuously tunable repetition rate. Leveraging all the comb lines of this device, RRMFC system can measure the distance at a high acquisition rate and a high accuracy by peak positioning in the time-domain interference pattern. By reducing the time of a single measurement, RRMFC system can carry out more independent distance measurements with limited sweeping bandwidth and speed of modulation frequency, thus increasing the maximum amount of distance information that can be provided by a single system over a limited time.

We envision the future development of RRMFC method and system on following aspects. First, by integrating RRMFC system along with the lasers[30-32], amplifiers[33-35], and OPAs[36] on a single thin film lithium niobate wafer, we could achieve a full solid-state LiDAR emitter with megahertz pixel rate. Second, considering that time-domain distance resolving method requires a higher signal-to-noise ratio, RRMFC can also resolve distance by calculating the frequency difference of the beat notes generated by all the sidebands in frequency domain, leading a higher precision and adaptability to various weather and illumination conditions. Thirdly, according to the difference of the measured distance, RRMFC can measure velocity with good acquisition rate and precision. Finally, RRMFC can be used as the primary ranging solution for high accuracy relative distance measurement methods, such as dual-comb ranging, to expand these methods' range with low response time and high stability.